\documentclass[]{spie}  

\usepackage{amsmath,amsfonts,amssymb}
\usepackage{graphicx}
\usepackage[colorlinks=true, allcolors=blue]{hyperref}

\title{Performance of the Upgraded VERITAS Stellar Intensity Interferometer (VSII)}

\author[a]{David B. Kieda}
\author[b]{the VERITAS Collaboration}
\affil[a]{Department of Physics and Astronomy, University of Utah, Salt Lake City, UT 84112 USA}
\affil[b]{https://veritas.sao.arizona.edu}

\authorinfo{Further author information: (Send correspondence to D.B.K.)\\D.B.K.: E-mail: dave.kieda@utah.edu, Telephone: 1 801 581 5220} 

\begin{document} 
\maketitle

\begin{abstract}
The VERITAS Imaging Air Cherenkov Telescope array (IACT) was augmented in 2019 with high-speed focal plane electronics to create a new Stellar Intensity Interferometry (SII) observational capability (VERITAS-SII, or VSII). VSII operates during bright moon periods, providing high angular resolution observations ( $<$ 1 mas) in the  B photometric band using idle telescope time. VSII has already demonstrated the ability to measure the diameters of two B stars at 416 nm (Bet CMa and Eps Ori) with   $< 5$\% accuracy using relatively short (5 hours) exposures\cite{SII2020}.

The VSII instrumentation was recently improved to increase instrumental sensitivity and observational efficiency. This paper describes the upgraded VSII instrumentation and documents the ongoing improvements in VSII sensitivity. The report describes VSII’s progress in  extending SII measurements to dimmer magnitude stars and improving the VSII angular diameter measurement resolution to better than 1\%.  
\end{abstract}


\section{THE VERITAS STELLAR INTENSITY INTERFEROMETER (VSII)}
\subsection{Overview}
The Stellar Intensity interferometry (SII) technique measures correlated fluctuations in light intensity between spatially separated telescopes.   The VERITAS Stellar Intensity Interferometer (VSII)  is currently the world's most sensitive SII astronomical observatory. VSII is implemented through an instrumentation augmentation of the existing VERITAS gamma-ray observatory \cite {Kieda2021}. VSII operates for 5-10 days/month during bright moon periods when the intensity of scattered moonlight severely limits VERITAS gamma-ray observations. This paper describes the VSII instrumentation, including recent upgrades and improvements in the VSII observatory.  This paper also updates the  comprehensive VSII survey of northern sky bright stars \cite{Kieda2021b} that is currently in progress. This survey has performed nearly 500 hours of observation on 48 different astronomical targets since December 2019. 

\subsection{VERITAS Observatory}
\begin{figure}[!htbp]
  \vspace{5mm}
  \centering
  \includegraphics[width=5.in]{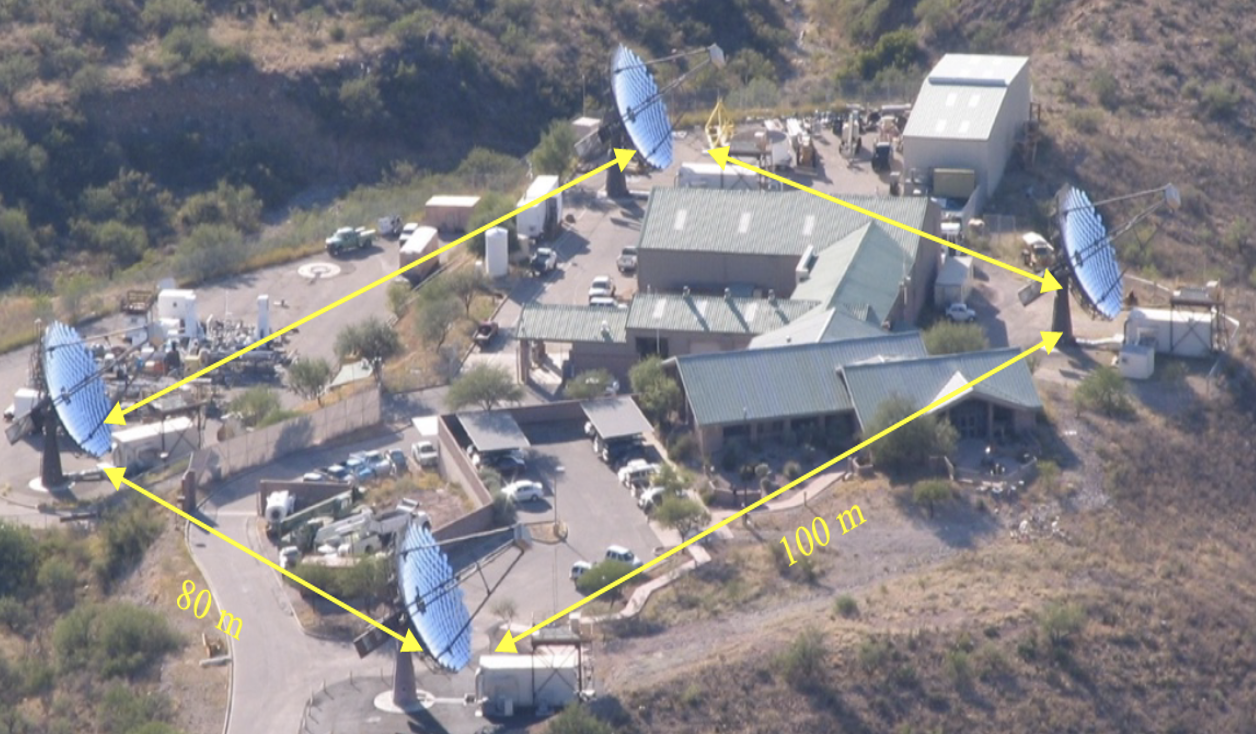}
  \caption{Baseline separations of the VERITAS array of 12-m diameter Gamma-ray telescopes (Amado, AZ USA) . The VERITAS Stellar Intensity Interferometer uses the VERITAS telescopes during bright moonlight periods when Very High Energy (VHE) gamma-ray observations are  not performed. }
  \label{Figure1}
 \end{figure}

The VSII Observatory uses the optical telescopes of the VERITAS gamma-ray observatory. The VERITAS Observatory  \cite{VERITAS} consists of an array of 4 IACTs located at an altitude of 1268 m.a.s.l at the F.L. Whipple Observatory (FLWO) near Amado, AZ. Each IACT contains an f/1.0 12 m-diameter segmented  Davies-Cotton reflector employing 345 hexagonal facets, resulting in 110 $m^2$ of light-collecting area. The lateral separation between nearest-neighbor IACT is approximately 80-120 m (Figure \ref{Figure1}). The Davies-Cotton reflector design creates approximately a 4 nanosecond spread to photons arriving at the telescope's focal plane. The combination of large telescope primary mirror area, fast optics, and 100+ meter telescope baselines enables VSII to provide sub-milliarcsecond angular resolution at short optical wavelengths (B/V bands)\cite{Matthews2020}

\subsection{VSII Focal Plane Instrumentation}
\subsubsection{Removable Focal Plane Plate}
\begin{figure}[!htbp]
  \vspace{5mm}
  \centering
  \includegraphics[width=3.in]{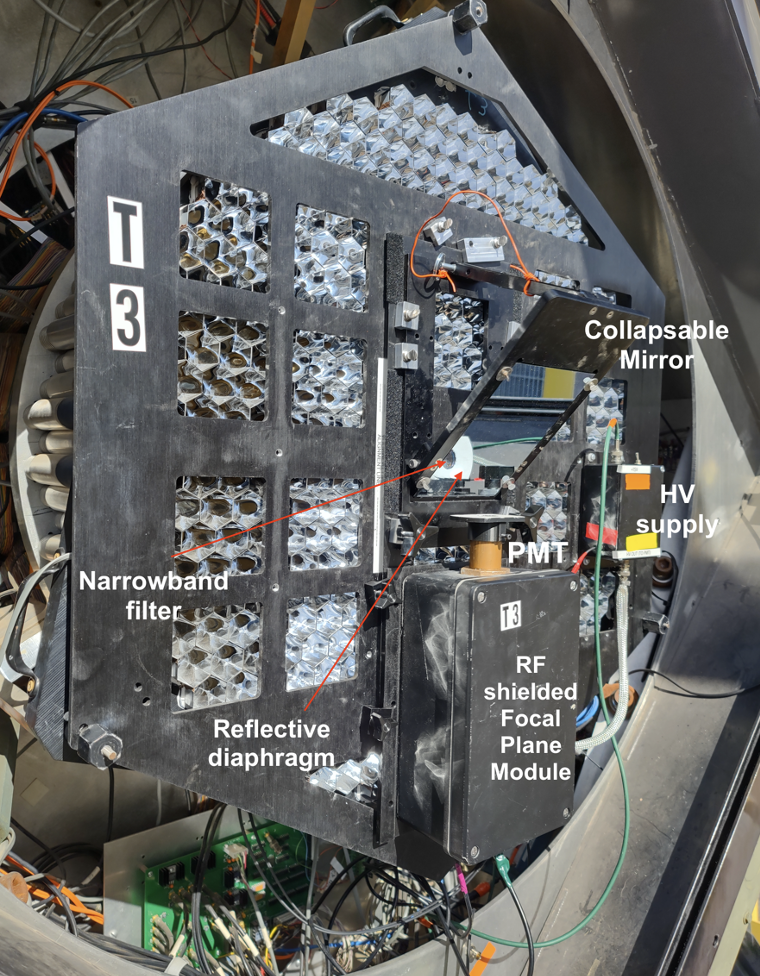}
  \caption{VSII removable Focal Plane Plate (FPP). The FPP contains a collapsible mirror, HV supply, Photomultiplier tube, RF shielded electronics box, and a narrowband optical filer. An FPP  is mounted in front of each VERITAS camera and is secured in place by hexagonal nuts tightened on three threaded rods.}
  \label{Figure2}
 \end{figure}

Each VERITAS telescope is instrumented with a removable VSII Focal Plane Plate (FPP)  that is securely mounted in front of the  499-pixel VERITAS camera (Figure \ref{Figure2}). 
  The removable FPPs  can be easily installed or removed in about 15 minutes per telescope by a single individual.  
The FPPs are designed to be securely mounted with no modifications or changes to the VERITAS cameras.
The FPP instrumentation employs a collapsible 45$^\circ$ mirror that reflects the starlight from the 12-m Davies-Cotton primary reflector onto an SII focal plane diaphragm that is oriented perpendicular to the telescope optical axis.
The collapsible mirror allows the VERITAS camera shutter to be closed during the daytime or bad weather.  This capability allows the SII FPPs to remain mounted on the VERITAS focal plane for the entire 5-10 day duration of the SII observing period.  

\subsubsection{Optical Instrumentation}
The SII focal plane employs  a Semrock FF01-420/5-2 narrowband optical filter (420 nm/5 nm width)  \cite{semrock}  located in a diaphragm mount that is located directly in front of a fast light sensor (Hamamatsu super-bialkali R10560 photomultiplier tube (PMT)). The diaphragm mount includes a central aperture for the Semrock filter. A flat oval-shaped white reflective screen  enables the observer to center the light from the primary mirror onto the Semrock filter. The SII FPP instrumentation plate scale is automatically matched to the VERITAS focal plane point spread function (PSF) as it uses the same size PMTs used for the VERITAS gamma-ray camera.

\subsubsection{Narrowband Optical Filter Effective Bandwidth}
The center bandpass wavelength of a narrowband optical interference filter will shift  to shorter wavelengths for light arriving at inclined angles.  The Semrock FF01-420 filter coating exhibits a high refractive index ($n = 2.38$) that reduces the wavelength shift for non-normal incident photons. A simulation was used to calculate the shift in optical bandpass wavelength for the VERITAS f/1.0 mirror optics by weighting the angular distribution of incoming photons by the relevant mirror area at each angle of incidence. The resulting incidence angle distribution was then convolved with the standard formula describing the shift in center bandpass wavelength with incidence angle\cite{semrock}. The resulting calculation indicates an effective bandwidth of $\delta \lambda \approx 13\ nm$  about a $\lambda_0 \approx 416\ nm$ center wavelength, with reduced light transmission compared to normal incidence.

\subsection{Signal Transport}
\subsubsection{Battery Powered High Voltage Supply}
 The PMT is powered by a custom battery-powered High Voltage  (HV) Power supply\cite{Cardon2019}. The HV supply battery uses two high-capacity (4400 ma-hr) TENERGY Li-Ion batteries permanently mounted in the VERITAS camera. The battery pack is recharged during the daytime.  The battery pack has sufficient capacity to provide HV for several nights of SII observations when fully charged. The HV setting is remotely controlled through a pulse width modulated  (PWM) signal supplied through a fiber optic interface. The fiber optic PWM signal is generated in the electronics trailer of each VERITAS telescope by an Ethernet-controlled Arduino Yun connected to a fiber optic transmitter.   
 
 Previously, the HV control system worked well in the lab, but occasionally would fail to control the HV module level adequately in the field, often requiring the observer to manually clean the fiber optic end and re-seat the fiber optic cable into the transmitter/receiver. This failure mode was eventually linked to the longer fiber optic cable being used on the telescopes than in  the lab, so  the LED driver light signal was marginally above the optoelectronic receiver's threshold.  In summer 2021, the LED fiber drivers were upgraded to an ultra-bright version (Industrial Fiber Optics IF-E96E) that has twice the light output of the original LED fiber drivers. After upgrading the LED driver, the HV control system has been operated for an entire year without encountering  additional failures. 
 
 \subsubsection{Preamplifier and Signal Cables}
The output of the PMT is amplified by a high-speed (200 MHz bandwidth) transimpedance ($2\times 10^4$ V/A) preamplifier (FEMTO HCA-200M-20K-C).  The distance between the PMT output and the input to the FEMTO preamplifier is kept as short as possible ($< $2 cm) to minimize the input capacitance load on the preamplifier input. This configuration maximizes the preamplifier bandwidth and decreases electronic noise.  The output from the FEMTO preamplifier drives a 45 m long double-shielded RG-223 coaxial cable. The cable is routed along the VERITAS telescope quadrapod arms and the optical support structure, and terminates at the SII data acquisition system, located in the telescope electronics trailer next to each VERITAS telescope. Special care is made to ensure that each SII focal plane instrumentation component is electrically isolated from any telescope ground to minimize ground-loop pickup of local RF noise. 
\subsection{Electronic Readout}
\subsubsection{Digitization}
\begin{figure}[!htbp]
  \vspace{5mm}
  \centering
  \includegraphics[width=6.7in]{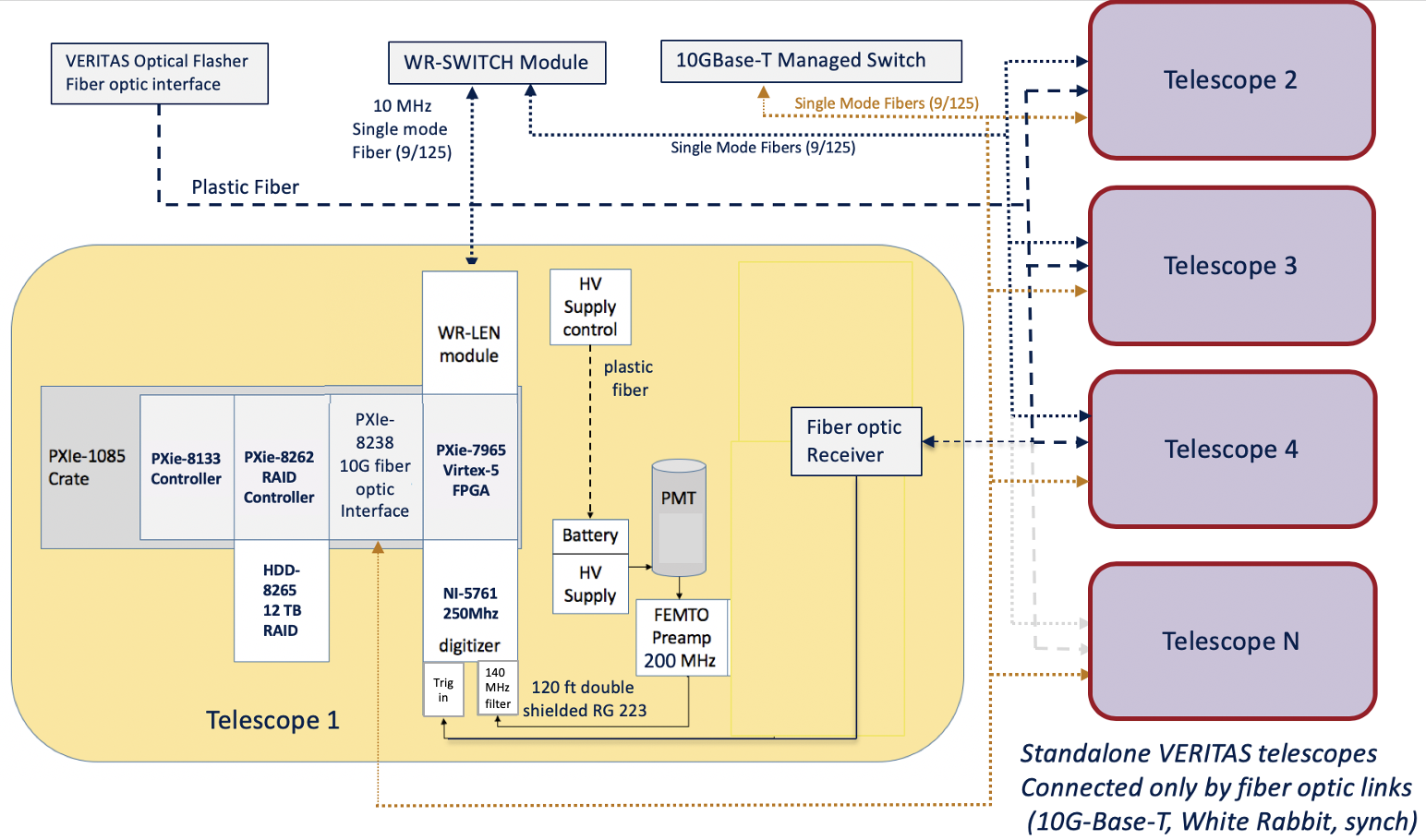}
  \caption{Schematics drawing of the 2019-era VSII electronics system. Each telescope contains a standalone VSII data acquisition system that operates independently. The timing of the DAQ clocks between telescopes is synchronized using a commercially available White Rabbit module from Seven Solutions.
  A separate fiber optic system is used to simultaneously trigger the start of the observing run at each telescope.}
  \label{Figure3b}
 \end{figure}

A standalone National Instruments (NI) data acquisition system,  using a PXI  Express (PXIe) crate with a high-speed backplane (4 GB/sec), is deployed at each VERITAS telescope to provide local data recording (Figure \ref{Figure3b}). Each SII data acquisition crate holds a high-speed PXIe controller (PXIe-8133) and a high-speed interface (PXIe-8262) connected to a high-speed (700 MB/sec) RAID disk (NI-8265). Each PMT signal is continuously digitized at 250 MHz by an NI FlexRio Module (NI-5761, 12-bit resolution,  DC-coupled). The 250 MHz digitization rate provides a sampling period that matches the 4 nsec time dispersion of the VERITAS optics. The digitized data from each PMT is truncated to 8 bits and merged into a single continuous data stream by a Virtex-5 FPGA processor. The data stream is transported across the high-speed PXIe backplane and recorded to the RAID disk. The VSII PXIe-based data acquisition uses commercially available modules, and the VSII data acquisition is controlled using LabView software. 

\subsubsection{RAID Storage}
VSII was initially configured to write continuously streaming data from the VSII observations to NI -8265 RAID storage systems. At each telescope, the 8265 RAID system, consisting of 12 independent 1 TB disks configured in a RAID0 configuration, supports continuous 700 MB/sec data streaming from the NI 5761 DAQ system digitizer. In the  2020-2021 observing season, the 1 TB drives were replaced by larger (2 TB) disks in each RAID system, raising the overall high-speed storage capability of each RAID drive to 24 TB.
In Spring 2021, all VSII crate controllers were upgraded to include high-speed USB 3.0 ports (PXIe-8135). This change allowed the mounting of additional low-cost RAID drive capacity (80 TB/telescope). These disks are used for archival storage, as their continuous write speed (~250 MB/sec)   is too slow to support reliable 250 MHz data streaming from the NI-5761 digitizers. 

In Fall 2022, two  crate controllers were upgraded to next-generation version (PXIe-8861) controllers with dual Thunderbolt-3 ports. The Thunderbolt 3 ports have sufficient bandwidth (3 GB/sec) to allow the streaming of the SII  DAQ data directly onto  very large capacity  (80 TB) RAID disks. This capability provides a low-cost expansion of available disk space for nightly observations, and sufficient disk space to run for 10+ nights without needing to process data. As of July 2022, each SII DAQ system has more than 150 TB of high-speed data storage available for observations.  At a sampling rate of 250 MHz (8-bit resolution), a one hour SII observation uses  865 GB of disk space. 

\subsubsection{Synchronization}
The intensity signals recorded between separated telescopes must be synchronized within a small fraction of the DAQ sampling period. If the DAQ clocks are not 100\% synchronized, the effective bandwidth of the measurement is degraded, resulting in a lower Signal-to-Noise Ratio (SNR). The distributed VSII data acquisition system incorporates synchronization of each telescope's DAQ FPGA clock using a commercially available Seven Solutions White Rabbit  clock synchronization system\cite{WR}. The White Rabbit timing system uses single-mode fiber (SMF) optic links to distribute a centrally generated 10 MHz clock to each SII DAQ system telescope with a $< 200\ psec$ RMS precision. 

\subsection{Two-telescope Correlations}
\subsubsection{High-Speed Data Transport}
An independent SII 10 GBase-T fiber optic data transfer network is used to provide high-speed network access to all four SII RAID data disks. This 10 Gb/sec bandwidth allows each VERITAS-SII DAQ system to cross-correlate with every other SII telescope data stream.  A 5/125 SMF  carries the  VSII high-speed datastream from each SII DAQ system to a central NetGear M4300-12X10 GBase-T managed Ethernet switch.  The NetGear switch isolates the SII-related data traffic from the primary VERITAS observatory network.  An independent 1 Gb/sec network interface is also established to each crate controller and HV control system in each telescope to allow access to these devices without competing for bandwidth with the main SII data stream. 

\subsubsection{FPGA-based correlators}
Cross-correlation of the data streams between individual telescopes is performed using a  pipelined algorithm implemented on a National Instruments FPGA card mounted in each VSII DAQ system.  The correlation system searches the directory structure of the previous night's SII observations, matches individual telescope observation files into pairs for correlation, and then stages a batch process to perform all identified correlation pairs. A  1-hour SII observation generates 865 GB of data per telescope, and the FPGA correlator requires approximately 1.5 hours to complete the correlation of a single telescope pair on the 1 hour observation.   The  VSII data transfer network has more than sufficient bandwidth to accommodate the simultaneous operation of multiple correlators on two independent DAQ systems.  

The main limitation in running multiple correlators in parallel is providing single-user access to the RAID disks. If two correlators are trying to access the same RAID disk on one telescope, the RAID will spend a substantial amount of time accessing  different files on the RAID disk. This will cause the correlation time to be significantly lengthened for both correlators. Nightly VSII operations have developed a more practical approach of using two RAID drives per telescope, alternating between writing to the two independent disks on sequential observing nights.   A single telescope can handle providing data to two independent correlator processes if the data is on separate disks, so in practice, up to four correlators may be run in parallel. 

\subsubsection{Processed File Size Reduction}
Each two-telescope correlation file is considerably smaller than the raw data file. A one-hour, two-telescope cross-correlation file is approximately 33 MB, compared to the combined 1.7 TB size of the two raw data files.  Once all two-telescope cross-correlations are completed, the correlation files are uploaded to the University of Utah Center for High Performance  Computing for archiving, distribution and analysis.  The majority of the raw data is discarded once the cross-correlations are completed. Secondary raw data files containing electronic calibrations and measurements of sky brightness are substantially smaller in size. Consequently,  it is possible to permanently retain them for correcting systematic effects for the final analysis. 
\subsubsection{Performance}
 For a one-hour observation using two independent correlators, approximately 4.5 hours of  VSII DAQ system processing  is required to process all six baseline pairs.  It has become routine practice to  process a fraction of the files the next day to assess the data quality from the previous night and process the remainder after the completion of the monthly SII observing period. In general, an entire week of nightly SII observations requires 7-10 days of computation to complete the correlations between every telescope pair.   

\subsection{RF Noise Reduction}
\begin{figure}[!htbp]
  \vspace{5mm}
  \centering
  \includegraphics[width=6.in]{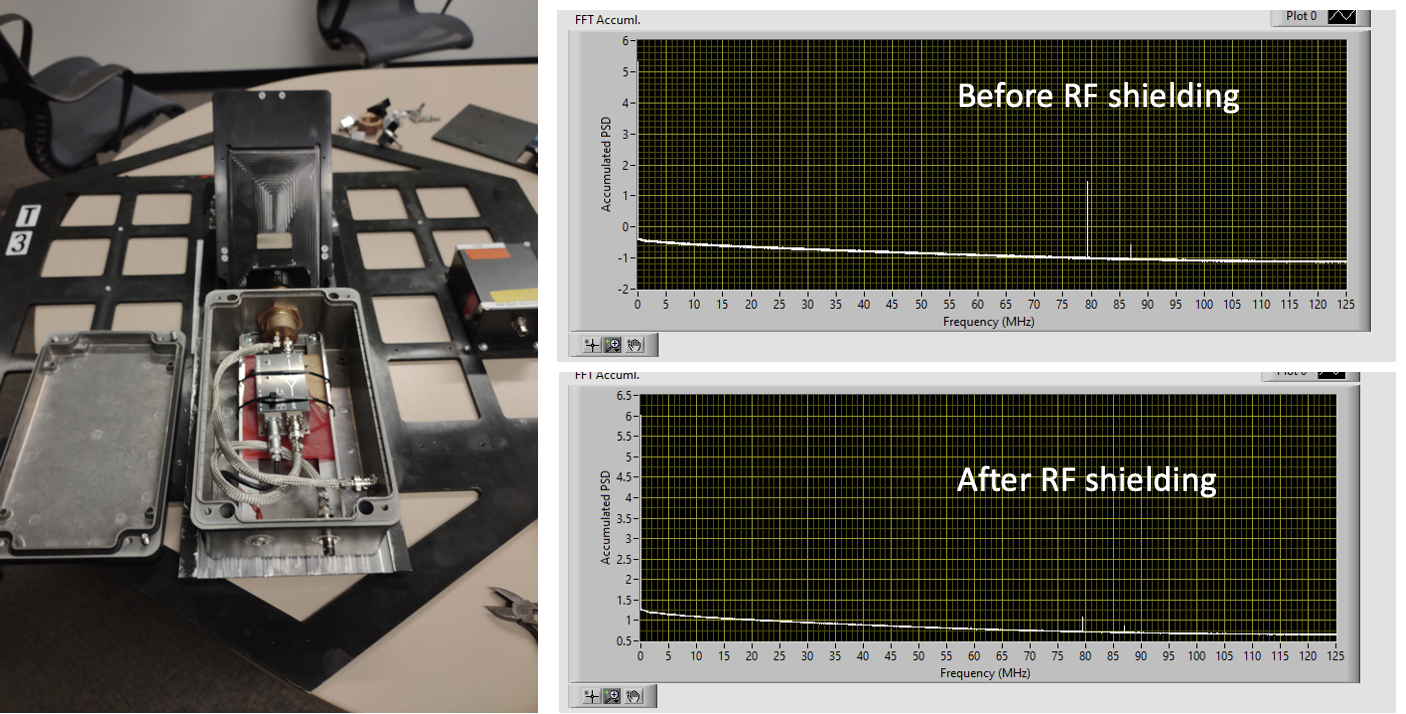}
  \caption{Effect of RF shielding of Telescope 3 Focal plane electronics. Left photo: RF shielded electronics enclosure for FPP electronics. Right Plots:  The noise spectrum is calculated using a standard Fast Fourier Transform  (FFT) on raw data from Telescope 3  during an observing run. Horizontal Axis: Frequency (MHz). Vertical axis: FFT power spectrum (arbitrary units). Upper plot: Noise spectrum before RF shielding. Bottom Plot: Noise spectrum after RF shielding.}
  \label{Figure4}
 \end{figure}

The photomultiplier tube (PMT) is encased in a tight-fitting brass tube to reduce RF pickup into the PMT dynodes.  However, FFT analysis of data from each SII PMT in December 2020  revealed the presence of both persistent and transient RF noise (Figure \ref{Figure4}, upper panel). Prominent RF pickup specific to the FLWO observatory site is observed at a measured frequency of 79 MHz.  This noise appears to be an alias of higher frequency RF noise undersampled at the 250 MHz digitization rate. Additional RF shielding was added to the FPPs in January 2021 to reduce RF pickup. The PMT and the preamplifier were fully encased in an RF-shielded enclosure (Focal Plane Module, or FPM). The FPM  reduced the RF pickup in each SII channel by a factor of 8 or more (Figure \ref{Figure4}, lower panel).   In addition, inline low-pass (140 MHz)  anti-aliasing filters (Mini-Circuits SLP-150+)  were installed on all VSII data acquisition systems in Summer 2021, and subsequent all observations have used the filters to reduce the RF pickup. 

After making these improvements, the residual RF  noise has been eliminated from any of the telescope traces, and the noise does not appear as a visible RF signal in any of the signal traces. There is still a residual level, however, which causes occasional `noise-assisted' flipping of the LSB of the digitizer  in the datastream. Although this RF effect is not visible in the raw datastream, it manifests as a low-level, persistent 79 MHz signal in the cross-correlations between telescopes. The exceptional power of the  cross-correlation analysis allows these random, correlated bit flips to be visible in the two-telescope correlations. Subsequent computational processing on the correlated data can effectively remove this residual component. In addition, we are installing (Summer 2022) triply shielded RG-223 signal cables;  these have demonstrated additional RF pickup attenuation by an additional 50\% (or more). 

\subsection{Zero Baseline Focal Plane Plate}
In June 2021, a specialized "zero baseline (ZB) beamsplitter"  FPP was constructed and deployed on VERITAS Telescope 1.  The ZB FPP adds a non-polarizing 50-50 beamsplitter after the narrowband filter to the optical beam path. Two independent FPMs mounted perpendicular to each other on the FPP read out the split optical beam. Special care is taken to ensure the FPMs do not share a common ground connection on the FPP to eliminate ground-loop noise pickup.   Each FPM uses independent batteries, HV supplies, signal cables, and DAQ systems in the trailer. The beamsplitter was constructed using commercially available PVC tubing combined with a commercial 25 x 35.36 x 1mm VIS, Elliptical Plate Beamsplitter (Edmund \#48-915). Successful commissioning of the ZB FPP was performed using SII observations of several stars in Summer/Fall 2021. 

 \begin{figure}[!htbp]
  \vspace{5mm}
  \centering
  \includegraphics[width=5.in]{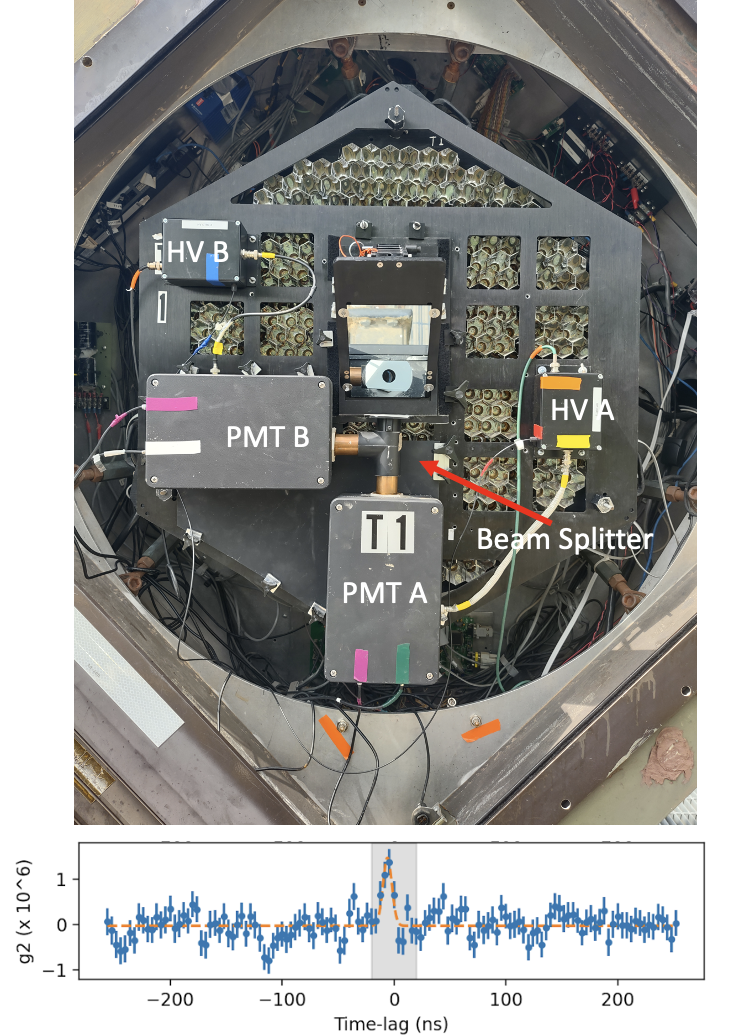}
  \caption{VSII Beamsplitter performance. Upper photo: VSII beamsplitter plate. The two PMTs (A, B) and their respective HV supplies are labeled. Bottom plot: Observed Two-Telescope Zero baseline correlation observed on Alp Lyr (Vega) October 16, 2021. The 30-minute observation was taken in modest moonlight conditions. Horizontal Axis: Time lag between T0 and T1 observation. Vertical axis: correlation $(g2-1) \times 10^6$.}
  \label{FigureZB}
 \end{figure}

The ZB FPP performance was evaluated in Fall 2021 by observing the bright A0V star Vega (Figure \ref{FigureZB}). One of the challenges of using the ZB FPP is that the light at each ZB PMT is reduced by a factor of 2 due to the beamsplitter, plus additional optical losses in the beamsplitter PVC pipe. The reduced photon flux at the beamsplitter PMTs also affects the correlation of Telescope 1 with the other telescopes in the VSII array, significantly reducing their sensitivity. To improve photon collection efficiency in the beamsplitter tube, in November 2021,  a front-surface mirror-polished aluminum 6061 foil (McMaster \#1655T1) was inserted into the PVC beamsplitter tube. The ZB FPP beamsplitter was used in observation in November-December 2021, and analysis of the observational data is currently underway.

\subsubsection{Summary of Hardware Configuration Changes} 

\begin{figure}[!htbp]
  \vspace{5mm}
  \centering
  \includegraphics[width=6.in]{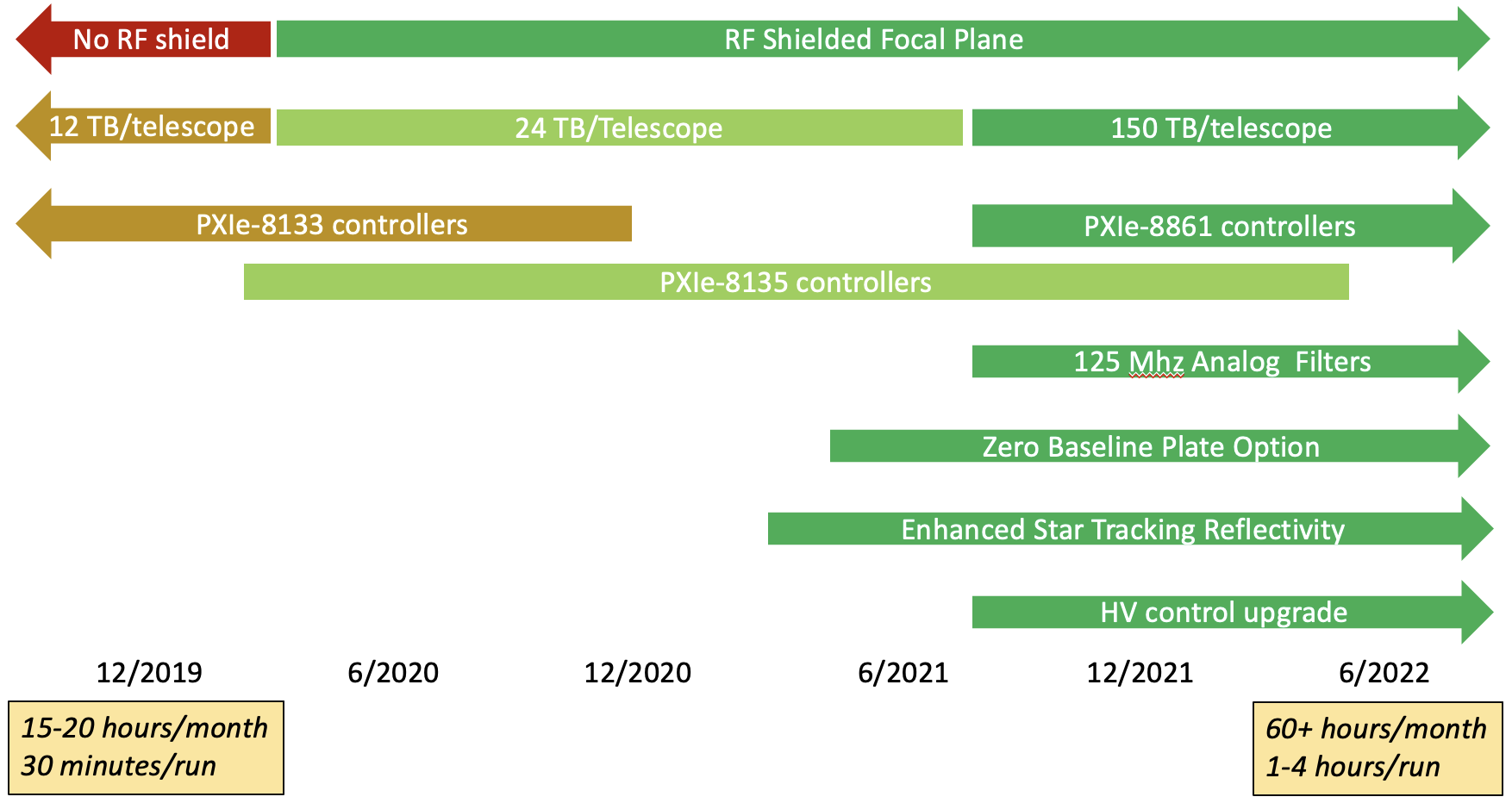}
  \caption{Timeline for various hardware improvements in the VSII observatory. VSII would record 15-20 hours of observations per month at the start of the survey period (December 2019). Presently, VSII system can record  more tha 60 hours of data during full moon periods when VERITAS does not perform VHE gamma-ray observations. Recent VSII system improvements have created the ability to perform multi-hour observations routinely. }
  \label{timeline}
 \end{figure}

Figure \ref{timeline} provides a summary of the hardware changes that have been made to the VSII instrumentation from the beginning of the Northern Sky SII survey (December 2019) until the time of this manuscript (July 2022). 

\section{VSII Observing}
\subsection{Overview}
 VSII has been in regular 4-telescope operations on a  lunar-monthly (full-moon/rear-full-moon period) cadence since December 2019, recording almost 500 hours of observations to date.  VSII  was not in operation from March 2020 through September 2020 due to the COVID-related shutdown of the Whipple Observatory.   The observing sequence has evolved to include nightly planning of observing constraints imposed by moon location and other  considerations (e.g., stellar magnitude, estimated stellar diameter, source elevation, weather, etc.)\cite{Kieda2021}.  
 
\subsection{Planning of Nightly Observations}
The suitability of a target for any given night of observation depends upon several factors. These factors include source-specific 
characteristics (stellar classification and magnitude, stellar diameter, single/binary/multiple star configuration),  observatory
characteristics (observatory latitude/longitude, telescope diameters, telescope separations, electronic noise, and optical efficiencies), and
astronomical considerations  (Source RA/Dec, number of hours observable on a given night, seasonal nighttime visibility of the source). 
VSII uses the public domain ASIIP software package \cite{davis,ASIIP}  to list and rank the suitability of potential sources for any given observation night.

ASIIP simulations are performed for a specified SII observing week during the year, and potential observation targets (drawn from several catalogs, including the JMMC catalog \cite{JMMC}) are ranked according to the estimated error in the determination of the stellar diameter.  For example, larger diameter  stars ($> 1 \ mas$) may not be resolved by the large VERITAS baselines, and the source radius can only be effectively constrained by low-elevations observations on a given night. In contrast, smaller stars ($< 0.5\ mas$) may be unresolved if the observations are constrained to  low elevations on a given night. 
At the beginning of the observing night, the simulated visibility curves for each target (e.g., Figure \ref{Figure1} ) are reviewed, and an observing sequence of 2-5 stellar observations is generated.  Then, the observing plan is sequenced to provide an appropriate set of observation hours and baselines for each target to result in a  quality measurement of each visibility curve.  

\subsection{Observing Sequence}
Preparations for Nightly SII observations begin about 1 hour before twilight. The observer opens the shutter to each VSII camera, and deploys a collapsible mirror  to a 45-degree position to reflect the starlight onto a photomultiplier tube (PMT) mounted transversely to the optical axis. The PMT High Voltage (HV) supply battery is re-connected after daytime charging, and the HV supply is powered up.  The Data Acquisition (DAQ)  system for each telescope  is then started, and short (10-30 second duration) calibration runs (with HV = 0V) are performed  with the telescopes parked in their stow positions. The pedestals of the data traces are recorded and used for removing the DC offset in each electronics channel. About 30 minutes before the start of the observation, a low HV level is turned on to verify each PMT is operational and that the telescopes are ready for beginning observations. The telescopes are then slewed to the first target of the night, and a VERITAS focal-plane CCD camera for each VERITAS telescope is used to micro-adjust each telescope's pointing, centering the point spread function of the starlight onto the narrowband filter.   Finally, the HV in each telescope is adjusted to provide a nominal DC current  in each PMT (typically 10 microamps for an $m_V = 2$ star).

Once the PMT currents are equalized, the telescopes are slewed $0.5^\circ$ off from the star's sky position, and a 1-minute OFF run is taken to measure the night sky background near the star's sky field. The telescope is then slewed to point at the target star, and a 30-180 minute continuous observation is performed (an ON run).  Each observation  run is triggered at every telescope by a central trigger pulse distributed to each DAQ system via a fiber optic system.   As the VERITAS telescopes track the target over 4-6 hours, the projected distance between the telescope combinations continuously changes with the movement of the star position in elevation and azimuth.  The measurement of the correlated fluctuations between two telescopes (interferometric visibility  $|V(r)|^2$) is therefore sampled over a range of projected distances $r$ during a single night’s
observations.

During the ON run, the starlight's point spread function (PSF) on the SII focal plane narrowband filter is tracked. If the PSF moves off the narrowband filter window, manual telescope tracking adjustments are performed to recenter the starlight on the light sensor. At the end of the ON run, another OFF run is taken with a $0.5^\circ$  offset. The  ON-OFF observing sequence is repeated throughout the night, as necessary.
Moving to a new observing target requires slewing to the desired target, making the manual tracking
adjustments to center the PSF onto the optical filter, setting the HV to equalize the PMT gains, and
then performing the nominal OFF-ON-OFF run sequence.

At the end of the observing night, the HV is turned off and the telescopes are brought to their stow positions.  The HV battery is re-connected to the battery charger, and the collapsible mirror is folded down. The camera shutter is then closed to protect the focal plane instrumentation during the daytime.
\subsection{Moon Angle Constraints}
 The narrowband Semrock  interference filter  (5 $nm$ bandpass, normal incidence)  \cite{semrock} substantially reduces the intensity of background light impinging on the VSII photomultiplier  tubes (PMTs).   This allows VSII observation to occur at all moon phases, including the full moon. In practice, the moonlight restricts the observability of  specific stars on a given night. The effects include:

\begin{enumerate}
\item Moonlight shining directly on the focal plane. Direct moonlight on the focal plane makes it difficult to view the location of the starlight PSF  on the interference filter aperture using the focal plane CCD camera.  The direct moonlight makes it challenging to visualize the tracking adjustments necessary to keep the starlight PSF centered on the narrowband filter.  To first order, this issue rules out any observations when the moon angle to the target source is  $> 90^\circ$. In practice, the VSII focal plane is slightly recessed into the VERITAS camera body, providing additional baffling against the moonlight. This baffling allows targets to be observed with moon angles $<95^\circ$. 

\item Moonlight is scattered by the atmosphere through both Rayleigh and Mie Scattering. The scattering functions  are peaked in the forward direction, creating a halo of scattered moonlight around the moon. SII observations are challenging when the target star is less than $30^\circ$  away from the moon's sky position. Moon  angles less than $30^\circ$  also make it challenging to make tracking adjustments. 

\item Atmospheric conditions can cause additional constraints on the observability of targets. Patchy clouds in the night sky can strongly scatter moonlight onto the focal plane at unanticipated angles, making it difficult to  perform reliable tracking or estimate night sky background biases in the visibility curves.  Low clouds on the horizon can restrict observation to higher elevations (longer baselines), making it difficult to sample the central peak of the visibility curve for  stars with diameters greater  than  1 mas. Alternatively, atmospheric conditions that restrict a given night's observations to  low elevations (short baselines) will result in smaller stars being unresolved.  

\item Observing stars at low elevation will noticeably dim their magnitude due to the increased atmospheric depth.  If moon angle constraints force  observations to occur only at low elevation angles, visible dimmer magnitude targets ( $m_V>2.5$ ) may not be feasible.

\item Dimmer stars ($m_V>3.0$ ) are difficult to track under most moonlight conditions, regardless of the moon angle. Therefore, the nightly observation plan must schedule time between astronomical twilight and moonrise/moonset to perform such observations. 

\item The VERITAS Telescopes are located adjacent to other telescopes (e.g., the prototype CTA Schwartzchild-Couder telescope \cite{pSCT}) and other buildings and equipment located at the FLWO basecamp. For moon elevations near rise or set (elevation $< 30^\circ$), the moonlight may offer serendipitous scatter from the parked SCT telescope primary or from other objects in the basecamp. The reflected moonlight can project an upward moonbeam into the night sky which occasionally crosses the path of the focal plane of a VSII telescope. This effect causes unanticipated illumination of the VSII focal plane. The scattering of moonlight off other telescopes or equipment is difficult to predict ahead of time. If it is observed, the observers assess whether it compromises the ability to perform reliable observations, If so, the observing schedule is changed to select a different star. 

 \item In addition, substantial sky background (compared to the target starlight) will bias the visibility curves, resulting in systematic errors in stellar radius measurement.
The night sky background light correction is most straightforward when the background light is less intense, and the background does not change significantly across short angular distances across the night sky.
\end{enumerate}

 \begin{figure}[!htbp]
  \vspace{5mm}
  \centering
  \includegraphics[width=6.7in]{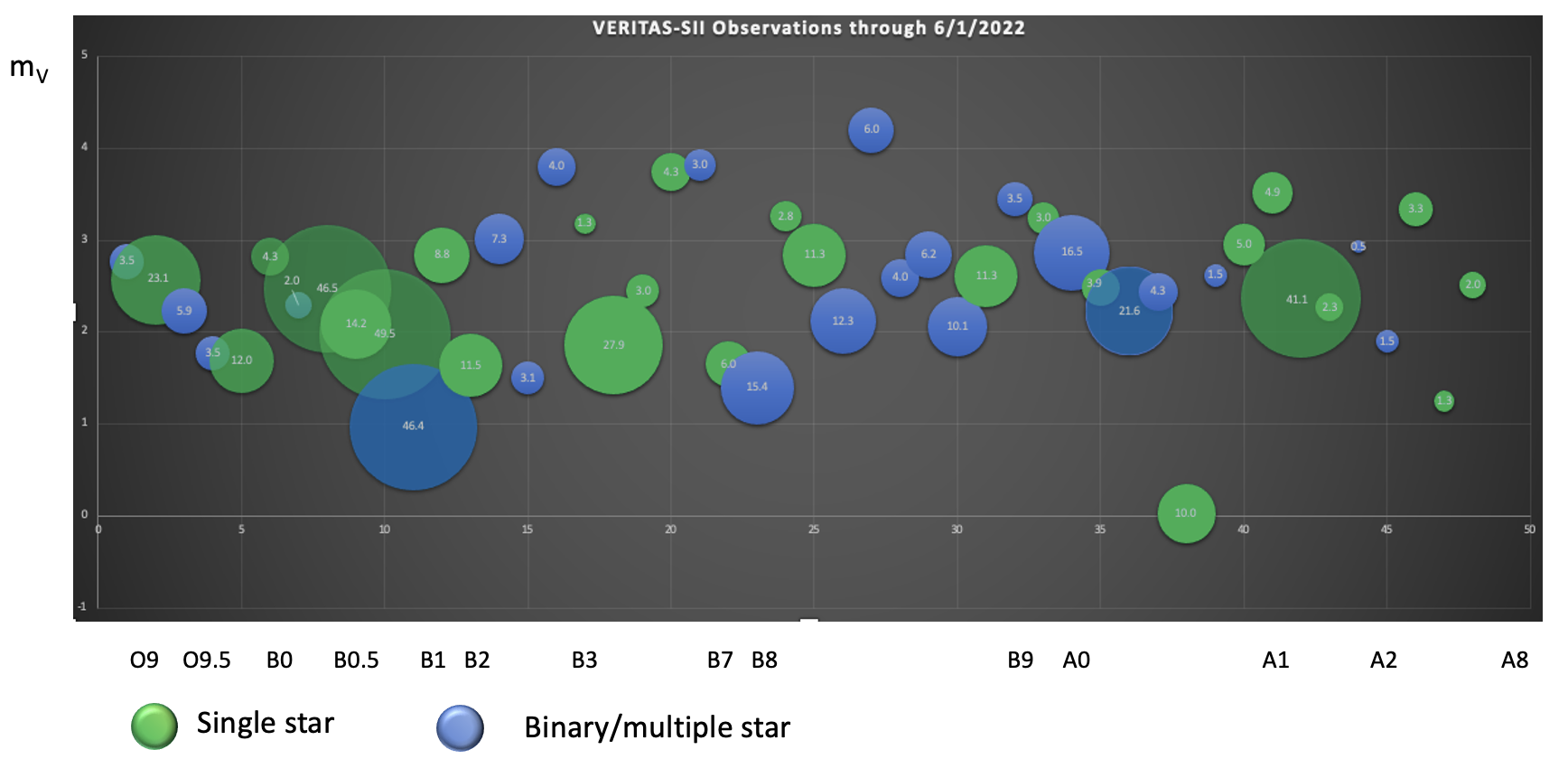}
  \caption{HR-diagram type summary of VSII survey Observations (12/2020-6/2022). Individual stellar observations are plotted as a function of stellar classification and visual magnitude.  Each circle represents a set of VSII observations. The area of the circle is proportional to the cumulative number of VSII observation hours. Green circles indicates single stars. Blue circles indicates binary or multi-star systems.  }
  \label{AllData}
 \end{figure}
 
 \subsection{VSII Observing Priorities}
 At the beginning of the observing week, the first target observed is chosen to
  be a bright SII reference (previously observed) star ($m_V < 2.2$) with
 $0.6\ mas < stellar \ radius < 0.85\  mas $   to verify the basic operational status of VSII.
 The star is chosen for an appropriate moon angle ($30^\circ < moon \ angle < 95^\circ$) and a strong 
 visibility signal with a 1 to 1.5-hour exposure.

 Priority for  subsequent observations is evaluated using the following priorities:
\begin{enumerate} 
\item $30^\circ < moon \ angle < 95^\circ$
\item Observation time  $> 1 $ hour
\item $0.4 \ mas < stellar \ diameter < 1.2 \ mas$
\item Quality of ASIIP constraint on the stellar diameter
\item Prefer $ m_V < 3 $
\item Prefer O, then B, then A stars
\item Previously unobserved  targets have a priority
\item Underexposed targets have a priority
\item Short period orbital binaries (e.g., Spica)  establish  a multi-day priority  to map out a visibility curve at different phases of the orbit. 
\item Unusual stellar characteristics (e.g., Cepheid, fast rotators, etc.) gain priority over `vanilla'-type singular stars.
\end{enumerate}

\section{VSII Observations 2019-2022}
From December 1, 2019 through June 1, 2022, VSII  has performed 496 hours of observations on 48 different targets. These targets include 26 single stars and 22 binary/multiple star systems.  Figure \ref{AllData} provides a "Hertzprung-Russel" (HR) type summary of the VSII observations. VSII  observations have been performed on stellar classifications ranging from O9 through A8, with an emphasis  on stellar classifications between O9 and B3. The stellar magnitudes observed  range between $+0.026 < m_V < +4.2$. Several targets have received exposures approaching 50 hours.

\subsection{Analysis of VSII Observations}
   Raw data from individual source observations are processed on-site into a "correllelogram"  using the pipelined two-telescope cross-correlation algorithm that  the FGPAs host  in the VSII Data Acquisition crates.  Once the correllelogram for each telescope pair is computed, each correllelogram is analyzed to extract the magnitude of the visibility at the specific telescope separation. Figure \ref{Figure4} illustrates the analysis steps of each two-telescope  correllelogram. This procedure is complicated by the presence of different levels of RF 79 MHz noise in the raw  data stream.
   
\begin{figure}[!htbp]
  \vspace{5mm}
  \centering
  \includegraphics[width=5.in]{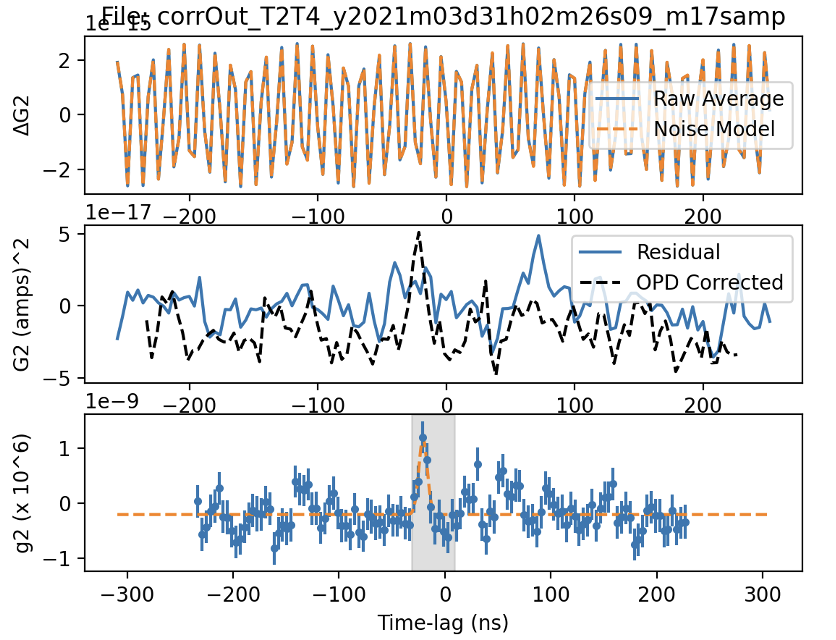}
  \caption{Analysis of a raw two-telescope correllelogram for the extraction of the visibility at a specific telescope separation. This observation of Zet Oph took place on March 31, 2021. The correllelogram measures the correlation between fluctuations in light intensities  recorded by VSII telescopes 3 and 4 as a function of the time lag between the photon arrival at each telescope. Upper panel: Fitting the raw correllelogram to remove RF noise pickup. Middle panel: Residual correlation signal before and after correction for optical path delay. Bottom panel: Fitting the Optical Path Delay (OPD) corrected correllelogram to a Gaussian model to extract the magnitude of the two-telescope correlation (visibility). Horizontal axis : time-lag between telescope data streams (nsec). Vertical axis: scaled interferometric visibility (scaling of units depends upon each panel)  }
  \label{Figure8}
 \end{figure}

   The first VSII observation took place before introducing the focal plane RF shielding and the anti-aliasing filters in 2020-2021. These observations had significantly larger RF noise levels  than the photocurrent signal stream.  The raw correllelogram is analyzed by fitting a noise model that includes a dominant 79 MHz component and several side frequencies. The weighting of each frequency component is iteratively adjusted to match the raw correllelogram data until the values converge on a satisfactory fit  (Figure \ref{Figure8}, upper panel). Next, the residual between the noise model and the raw data is calculated, and the residual data is corrected for the changing optical path delay during the  observation using the known projected distance (OPD) difference between the two telescopes to the target   (Figure \ref{Figure8}, middle panel). Finally a Gaussian fit is used to extract the peak of the visibility curve at the expected time lag between telescopes (Figure \ref{Figure8}, bottom panel). 
   
\begin{figure}[!htbp]
  \vspace{5mm}
  \centering
  \includegraphics[width=6.in]{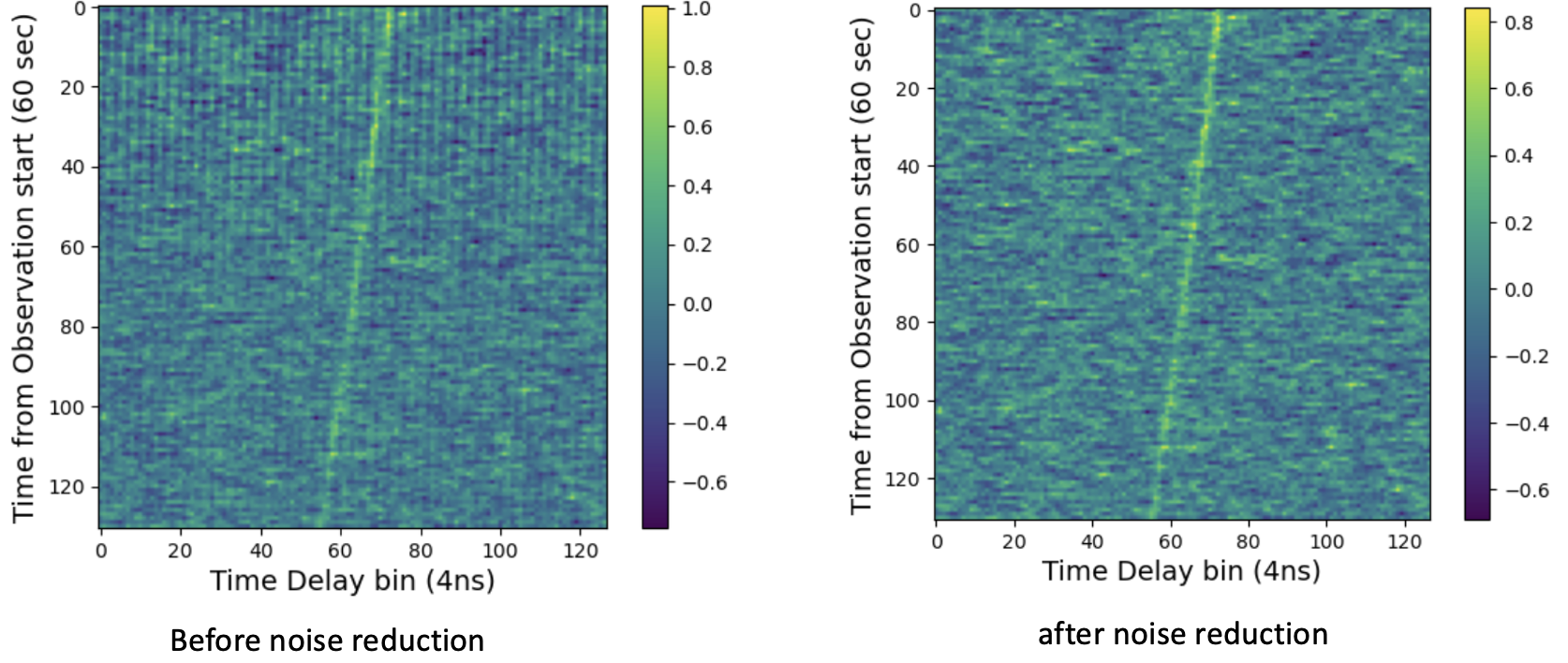}
  \caption{2D heatmap of a single observing run correllelogram before and after RF noise subtraction. The heatmap illustrates the time evolution of the T3-T4 correllelogram (horizontal axis, the color scale is the g2 correlation magnitude) as a function of time from the start of the  observing run. Yellow vertical line: evolution of the g2 peak through the observing run. Left panel: Raw correllelogram data, including RF noise. The  RF noise pickup is  seen as the correlated vertical ripples in the background (blue) region. Right panel: Residual correlation signal after applying RF noise rejection. The residual noise fluctuations in the background region are Gaussian distributed;  this noise can be used to estimate the limiting magnitude of the VSII observatory.  The observations were taken on December 20, 2021 on the star Eta Uma.   }
  \label{heatmap}
 \end{figure}

   Subsequent VSII observations incorporated RF shielded focal planes (Spring-2020) and 140 MHz anti-aliasing filters (Summer 2021).   These observations had RF pickup noise which is a small fraction of the photon stream signal. For these observations, an FFT is applied to the correllelogram data to identify the RF noise frequencies, and a combination of a Fourier Series model and machine learning tools is be applied to eliminate the RF noise contribution. Figure \ref{heatmap} demonstrates the reduction of RF noise in more recent VSII observations. The left heatmap shows the correlation signal on the T3-T4 correllelogram measured on Eta Uma in December 2021. The g2 coherence signal is seen as the bright yellow vertical line , and the RF noise contamination is seen as the striped patterns in the background fluctuations to the left and the right of the g2 signal. The plot on the right shows the same data after RF noise subtraction; the correllelogram is free from any noticeable level of RF noise.  Identifiable g2 peaks have been identified for VSII observations of stars as dim as $m_V = +3.75$.
   
   \subsection{Visibility Curves and Measurement of Stellar Diameters}
   
   A visibility curve is  calculated using the above-measured visibility peaks and known telescope separations for every telescope pair in the observation. The final visibility curve must be corrected for the presence of night sky background light. After this correction, a suitable stellar diameter model is  fitted to the visibility curve to accurately measure the stellar diameter \cite{SII2020,Matthews2020} at an effective  wavelength $\lambda = 416\ nm$. We perform our preliminary analysis fits using a Uniform Disk (UD) model to the visibility $|\gamma(d)|^2$ as a function of the projected telescope separation $d$ : $$|\gamma (d)|^2=g^{(2)}(\tau=0,d)-1$$
   $$|\gamma(d)|= 2 \frac{J_1(\pi\theta d/\lambda)} { \pi \theta d/\lambda}$$. 
   
\begin{figure}[!htbp]
  \vspace{5mm}
  \centering
  \includegraphics[width=6.7in]{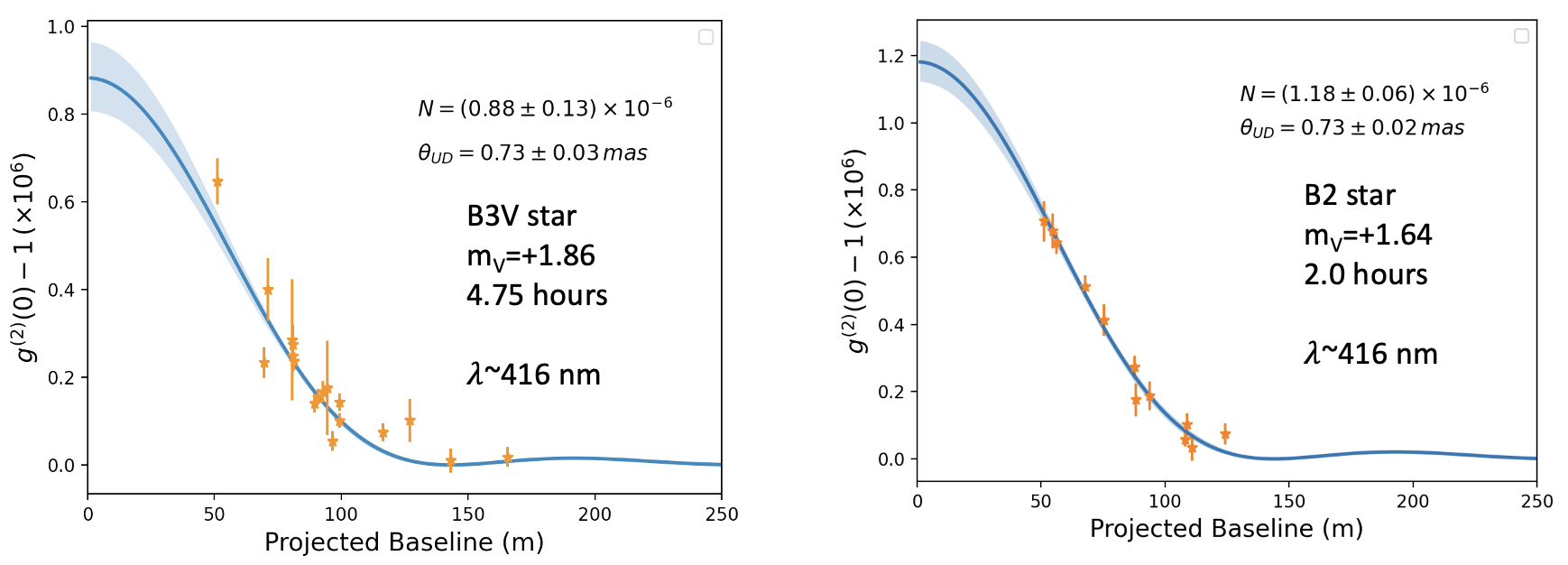}
  \caption{VSII observations of visibility curves for two bright B stars.  The analysis is preliminary as they have been uncorrected for the night sky background.  Uniform Disk models have been fitted to the visibility curves demonstrating the ability to measure the UD diameter to an accuracy of 3-4\% with short (2-5 hour) observations.  }
  \label{TwoSII}
 \end{figure}

   Two preliminary visibility curves obtained by recent VSII observations  are shown in Figure \ref{TwoSII}. With the present UD analysis, VSII  routinely provided measurements of stellar diameters with 3-4\% resolution for bright stars ($m_V = 1.5-2$) with short (2-5 hour) exposures. 
   Improvements in the VSII telescope hardware, analysis, and calibration are projected to improve the resolution in the future to 1\% or better. These improvements are the main focus of VSII analysis during 2022-2023. 
 
\section{Acknowledgements}
Both observations and analysis of data for the VERITAS Observatory are supported by grants from the US Department of Energy Office of Science, the US National Science Foundation, the Smithsonian Institution, and NSERC in Canada. The authors gratefully acknowledge support under NSF Grant \#AST 1806262 for the fabrication and commissioning of the VERITAS-SII instrumentation. We acknowledge the excellent work of the technical support staff at the Fred Lawrence Whipple Observatory and the collaborating institutions in the construction and operation of the instrument.


\end{document}